\documentclass[pra,twocolumn,showpacs,amsfonts,amsmath,floatfix]{revtex4}

\usepackage{graphicx}
\usepackage{epsfig}
\usepackage{epstopdf}
\usepackage[usenames,dvipsnames]{xcolor}
\usepackage{braket}
\usepackage{hyperref}
\usepackage{amsthm}
\usepackage{enumitem}

\newcommand{\lt}{\left(}
\newcommand{\rt}{\right)}

\newcommand{\be}{\begin{equation}}
\newcommand{\ee}{\end{equation}}
\newcommand{\ba}{\begin{eqnarray}}
\newcommand{\ea}{\end{eqnarray}}
\newcommand{\fr}{\frac}
\newcommand{\nn}{\nonumber}

\newcommand{\R}{\mathbb{R}}
\newcommand{\bc}{\begin{center}}
\newcommand{\ec}{\end{center}}
\newcommand{\beq}{\begin{equation}}
\newcommand{\eeq}{\end{equation}}
\newcommand{\beqq}{\begin{equation*}}
\newcommand{\eeqq}{\end{equation*}}
\newcommand{\beqa}{\begin{align}}
\newcommand{\eeqa}{\end{align}}
\newcommand{\barr}{\begin{array}}
\newcommand{\earr}{\end{array}}
\newcommand{\bi}{\begin{itemize}}
\newcommand{\ei}{\end{itemize}}

\newcommand{\C}{\mathbb{C}}

\input{Qcircuit}

\newtheorem{theo}{Theorem}
\newtheorem{conj}{Conjecture}
\newtheorem{pb}{Problem}

\begin{document}

\title{Continuous-Variable Sampling from Photon-Added or Photon-Subtracted Squeezed States}

\author{U. Chabaud$^1$}
\email{ulysse.chabaud@gmail.com} 
\author{T. Douce$^1$}
\author{D. Markham$^1$}
\author{P. van Loock$^2$}
\author{E. Kashefi$^{1,3}$}
\author{G. Ferrini$^2$}
\email{giulia.ferrini@gmail.com} 
\address{$^1$ Laboratoire d'Informatique de Paris 6, CNRS, UPMC - Sorbonne Universit\'es, 4 place Jussieu, 75005 Paris}
\address{$^2$ Institute of Physics, Johannes-Gutenberg Universit{\"a}t Mainz, Staudingerweg 7, 55128 Mainz, Germany}
\address{$^3$ School of Informatics, University of Edinburgh, 10 Crichton Street, Edinburgh, EH8 9AB}
\date{\today}

%-------------------------------------------------------------------------------------------------------------------------------------------------------------------------
\begin{abstract}

We introduce a new family of quantum circuits in continuous variables and we show that, relying on the widely accepted conjecture that the polynomial hierarchy of complexity classes does not collapse, their output probability distribution cannot be efficiently simulated by a classical computer. These circuits are composed of input photon-subtracted (or photon-added) squeezed states, passive linear optics evolution, and eight-port homodyne detection. We address the proof of hardness for the exact probability distribution of these quantum circuits by exploiting mappings onto different architectures of sub-universal quantum computers. We obtain both a worst-case and an average-case hardness result. 
Hardness of Boson Sampling with eight-port homodyne detection is obtained as the zero squeezing limit of our model. 
We conclude with a discussion on the relevance and interest of the present model in connection to experimental applications and classical simulations.

\end{abstract}
%-------------------------------------------------------------------------------------------------------------------------------------------------------------------------
\maketitle 

%-------------------------------------------------------------------------------------------------------------------------------------------------------------------------
\section{Introduction}

In the recent years we have witnessed an increasing interest in quantum circuits that define sub-universal models of quantum computation~\cite{Aaronson2013, Bremmer2010, Bremner2015,Fahri2016,Morimae2014}. These models lie somewhere in-between classical and universal quantum computing, in the sense that, although not possessing the full computational power of a universal quantum computer, they allow for the outperformance of classical computational capabilities with respect to specific problems. Beyond their conceptual relevance, the reason for this interest is that these models require less experimental resources than universal quantum computers do. Therefore, they may allow next future experimental demonstration of {\it quantum advantage}, i.e. the predicted speed-up of quantum devices over classical ones for some computational tasks. 

These models are often associated with sampling problems for which the task is to draw random numbers according to a specific probability distribution. Some of these probability distributions are likely to be hard to sample for classical computers, assuming widely accepted conjectures in computer science, for example with the celebrated Boson Sampling~\cite{Aaronson2013}.

In parallel, Continuous-Variable (CV) systems are being recognized as a promising alternative to the use of qubits, as they allow for the deterministic generation of unprecedented large quantum states, of up to one-million elementary systems~\cite{yokoyama2013optical, Yoshikawa2016}, and also offer detection techniques, such as homodyne, with high efficiency and reliability.

Any given CV quantum circuit is defined by (i) a specific input state, (ii) a unitary evolution and (iii) measurements. An important theorem~\cite{Bartlett2002, Mari2012} states that if all these elements are described by positive Wigner functions, then there exists a classical algorithm able to efficiently simulate this circuit. Hence, including a negative Wigner function element is mandatory in order to design a CV sub-universal quantum circuit that cannot be efficiently simulated by a classical device. By virtue of the Hudson theorem~\cite{soto1983wigner}, this necessarily corresponds to the use of non-Gaussian resources.

Therefore, if one aims at minimal extensions of Gaussian models, three different families of non trivial sub-universal quantum circuits can be defined, depending on whether the element yielding the Wigner function negativity is provided by the input state, the unitary evolution, or the measurement. Although Wigner negativity allows stepping outside the range of applicability of the theorem in Ref.~\cite{Mari2012},  it is by itself  not sufficient to imply classical hardness~\cite{Rahimi-Keshari2016}. 
The classical hardness of circuits of the latter kind, corresponding to Gaussian Boson Sampling (GBS), was proven in Ref.~\cite{Lund2014, Hamilton2016}. 
These circuits are composed of input squeezed states, passive linear optics evolution, and photon counters. Circuits of the second kind are for instance related to the CV implementation of Instantaneous Quantum Computing --  another sub-universal model, where input states and measurements are Gaussian, while the evolution contains non-Gaussian gates~\cite{Douce2017, Douce2018}. First definitions of the former class of CV circuits, i.e. that display non-Gaussian input state and Gaussian operations and measurements, have been very recently considered~\cite{Chakhmakhchyan2017, Lund2017}. However, the measurement considered in~\cite{Lund2017} is effectively a non-Gaussian measurement~\footnote{In principle, the model in Ref.~\cite{Lund2017} could be mapped onto a model with non-Gaussian input states and Gaussian measurement by moving to the input the single photon states that are used to implement the measurement, and by considering a larger unitary transformation with a block-structure corresponding to implementing beam-splitters between these modes and the modes to be measured. The  equivalence of the two models however only follows through in the worst-case proof scenario.}. Furthermore, the non-Gaussian input state considered in both references is a collection of single photons, that is still reminiscent of the standard Boson Sampling approach~\footnote{We also mention that in Refs.~\cite{Seshadreesan2015, Olson2015, Rohde2015} CV versions of the standard Boson Sampling are considered, where however both input and measurements are non-Gaussian: the measurement is still photon counting, while the input state is a CV non-Gaussian state, either a photon-added or subtracted squeezed state~\cite{Seshadreesan2015, Olson2015}, or a cat state~\cite{Rohde2015}.}. 

In this work, we define a new family of quantum circuits that take non-Gaussian input states and use Gaussian evolution and measurement. The circuits family that we consider has a further ``CV flavor" with respect to Refs.~\cite{Chakhmakhchyan2017, Lund2017}, in that the non-Gaussian input states are single photon-subtracted (or single photon-added) squeezed states, and the measurement is Gaussian, namely eight-port homodyne detection~\cite{Leonhardt-essential}~\footnote{Some authors, e.g. in Ref.~\cite{Chakhmakhchyan2017}, refer to this kind of detection as to heterodyne detection. We want to avoid this terminology here, as the expression ``heterodyne" is also often used to indicate a common detection technique in classical electronics.}. This model therefore is analog to the Photon-Added or photon-Subtracted Squeezed Vacuum (PASSV) sampling model of Ref.~\cite{Olson2015} but with eight-port homodyne detection replacing photon counting. For this reason, we may refer to our model as to PASSV with CV Sampling. For brevity, however, we will use the acronym CVS in the remainder of our paper.
%By virtue of the CV character of  both input states and output measurements, we refer to it as Continuous-Variable Sampling (CVS).
%
This architecture is inspired by recent experiments performed at Laboratoire Kastler Brossel (LKB), where mode-selective single photon subtraction from a collection of multi-mode squeezed states has been recently demonstrated~\cite{Ra2017}, and where simultaneous detection of all the optical modes can also be implemented by means of multi-pixel homodyne detection~\cite{Beck00, ferrini2013compact}.

%In order to establish a proof of hardness, we exploit the idea of time-reversal, i.e. the time symmetry of the Born rule, that was already used in Ref.~\cite{Chakhmakhchyan2017} and proposed in~\cite{Hamilton2016}. 
Importantly, in contrast to Refs.~\cite{Chakhmakhchyan2017,  Lund2017}, we are able to construct our proof as an ``average-case" statement: we show that if two conjectures hold true then the underlying problem is hard with high probability. More specifically, we exhibit a construction that allows one to draw at random hard to sample circuits with high probability.

%Due to the use of time-reversal, as in Ref.~\cite{Chakhmakhchyan2017} our approach requires introducing auxiliary family of circuits, that possess a structure similar to the ones addressed in Ref.~\cite{Hamilton2016}, and establishing a link between their output probability distribution and the hafnian of a matrix. In turn, the latter matrix is related to output covariance matrix of this auxiliary architecture.

In Sec.~\ref{se:model-definition} we define the model we are interested in, focusing on photon subtraction at the input, and in Sec.~\ref{se-proof} we outline the proof of hardness of the corresponding output probability distribution, detailing the mappings needed to establish it. We give the detail of this proof in Sec. \ref{se:Hardness-proof}. In Sec. \ref{se:remarks} we show that the proof can be easily extended to the case of input photon-added squeezed states, and we discuss other relevant extensions, such as the zero squeezing limit, and the choice of the measurement quadrature. Sec.~\ref{se:Exp} is dedicated to discussing possible experimental implementations of our model. Conclusions and perspectives are presented in Sec.~\ref{se:conclusions}.

%-------------------------------------------------------------------------------------------------------------------------------------------------------------------------
\section{Definition of the circuit family}
\label{se:model-definition}

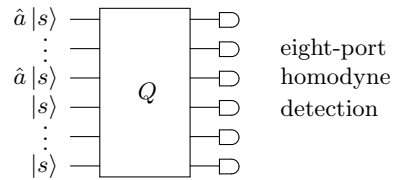
\begin{figure}
\begin{equation*}
\Qcircuit @C=1.2em @R=0.3em {
    \lstick{\hat{a}\ket{s}} & \multigate{5}{\phantom{aa}Q\phantom{al}} &\measureD{} & \\
     \overset { \vdots  }{\phantom{a}}\phantom{blabi} & \ghost{\phantom{aa}Q\phantom{al}} & \measureD{} & \rstick{\text{eight-port}}\\
    \lstick{\hat{a}\ket{s}} & \ghost{\phantom{aa}Q\phantom{al}} &\measureD{} & \rstick{\text{homodyne}}\\
    \lstick{\ket{s}} & \ghost{\phantom{aa} Q\phantom{al}} &\measureD{} & \rstick{\text{detection}} \\
     \overset { \vdots  }{\phantom{a}}\phantom{blabi} & \ghost{\phantom{aa}Q\phantom{al}} & \measureD{} & \\
    \lstick{\ket{s}} & \ghost{\phantom{aa}Q\phantom{al}} & \measureD{} &
}
\end{equation*}
\caption{A representation of a CVS circuit. In input are vacuum squeezed states and photon-subtracted vacuum squeezed states. The passive linear optics evolution is associated with the unitary matrix $Q$ defined in Eq.(\ref{eq:definition-Q}). Measurement is performed by eight-port homodyne detection.}
\label{fig:CVS}
\end{figure}

%By virtue of the CV character of both input states and output measurements, we call Continuous-Variable Sampling (CVS) the model which hardness we want to prove (Fig.~\ref{fig:CVS}). 
CVS circuits, for which classical hardness will be proven in this paper, are defined as follows. Let $M$ be the total number of optical modes. We define the squeezing operator with squeezing parameter $s$ as $\hat{S}(s)=e^{-\frac{i}{2}(\ln{s})(\hat{q}\hat{p}+\hat{p}\hat{q})}$. The input is made of vacuum squeezed states $\ket{s} \equiv \hat{S}(s)\ket{0}$, and the squeezing is uniform over all the modes. We adopt the convention $[\hat q, \hat p] = i$ for the quadratures commutation relation, i.e. $\Delta_0 = 1/2$ for the vacuum fluctuations. Given the corresponding action on the quadratures
$
\begin{pmatrix} \hat q  \\ \hat p \end{pmatrix} \rightarrow 
\begin{pmatrix} s & 0 \\ 0 & 1/s \end{pmatrix} \begin{pmatrix} \hat q  \\ \hat p \end{pmatrix},
$
$s > 1$ results in $ p$-squeezing while $s < 1$ in $q$-squeezing.

The first $m$ modes are single photon-subtracted squeezed states denoted by $ \hat a \ket{s}$. The remaining $M-m$ modes are just squeezed states $\ket s$. We assume that the squeezing parameter $s$ is  constant and does not depend on the number of modes $m$.
We require that $m$ is even and that $M\geq2m$. %\textcolor{red}{This will be used in Lemma [ADD LEMMA]}.
%The requirement that $m$ is even, instead, comes from necessity of consistency with the auxiliary TR-CVS model that we introduce. In the latter model, namely, only even number of photons can be detected, given that at the input are squeezed states. 
% 

The input modes undergo a passive linear evolution that is described by the unitary matrix $Q$ that belongs to the set of matrices of the form
\begin{equation}
\label{eq:definition-Q}
Q=\Theta e^{-i\phi\Sigma}
\end{equation}
with $\phi \in \R, \Theta\in \mathcal{O}(M)$ and $\Sigma\in\mathcal O_S(M)$, i.e., $\Theta$ is a real orthogonal matrix and $\Sigma$ is a real symmetric orthogonal matrix, and hence satisfies $\Sigma^2= 1$.
This choice of matrices will allow us to derive a convenient expression for the output probability distribution of CVS circuits in Sec. \ref{se:Hardness-proof}.

Finally, the mode quadratures are measured by eight-port homodyne detection, i.e. by projecting the output states onto displaced squeezed states $\ket{\alpha_i,r} =  \hat D(\alpha_i) \hat S(r) \ket{0}$. The arbitrary parameter $r >0$ is constant with the number of modes and $\alpha_i=  \sqrt{(1+r^2)/2}(q_i + i p_i/r)$ corresponds to the displacement value of the $i^{th}$ mode. 
$q_i$ and $p_i$ are the measured outcomes at the (distinct) output modes of the eight-port homodyne detector. $\hat D(\alpha)$ is the displacement operator $\hat D(\alpha)= e^{\alpha \hat a^{\dagger}- \alpha^* \hat a}$. This modelization is presented in detail  in  Appendix~\ref{app:heterodyne} (see also Ref.\cite{Leonhardt-essential})~\footnote{Note the difference in the notations for the squeezing operator $\hat S(r)$ in our work and in Ref.~\cite{Leonhardt-essential}.}.
%\textcolor{red}{[WRITE CORRECT EXPRESSION ALPHA, SEE APPENDIX]} The modelization is presented in.
%This measurement technique consists in measuring the $ p$ and $ q$ quadratures respectively of the two output modes emerging from a $XXX(r)$reflectivity beam splitter, that mixes the input signal with vacuum. 
The limit of perfect or zero reflectivity (corresponding to projection onto infinitely squeezed states) yields back the case of standard homodyne detection.
%The heterodyne detection actually corresponds to the measurement of the rescaled quadratures $r{q}_i$ and ${p}_i/r$ for the $i^{th}$ mode. 
As already considered in Refs.~\cite{Douce2017, Chakhmakhchyan2017}, in order to give meaning to the obtained probability distribution, that must be defined on discrete values, we assume a finite binning of size $\eta$ for the output probability density of CVS circuits~\cite{Paris2003}.
This allows for the definition of a set of indices $\bar { b }=(b^{(q)}_1,\ldots,b^{(q)}_M,b^{(p)}_1,\ldots,b^{(p)}_M)$ that corresponds to bins for the ${q}$ and ${p}$ quadratures. We denote  $\text{Pr}_{\rm CVS}^\eta(\bar { b }| \bar { n })$ as the discrete probability that the $i^{th}$-mode measurement outcome $(x_i,x_{i+M})=(q_i,p_i)$ falls into the boxes $(B^{(q)}_i,B^{(p)}_i)=[({ b }^{(q)}_{ i }-\frac12)\eta,(b^{(q)}_i+\frac12)\eta],[({ b }^{(p)}_{ i }-\frac12)\eta,(b^{(p)}_i+\frac12)\eta]$ for all $i\in\{1,\ldots,M\}$, given the input $\bar { n }=(n_1,\ldots,n_M)\in\{0,1\}^M$, where the ones correspond to photon-subtracted squeezed states and the zeros to squeezed states. Note that this probability distribution is related to the real-valued probability density associated with CVS circuits, $\text{Pr}_{\text{CVS}}(\bar x \vert\bar { n } ) $, by
\begin{equation}
\label{distribution}
\text{Pr}^\eta_{\text{CVS}}(\bar { b }| \bar { n })\equiv \prod _{ i=1 }^{ M }{ \left[ \int _{ B^{(q)}_i  }\int _{B^{(p)}_i  }{ \mathrm d{ q }_{ i }\mathrm d{ p }_{ i }\text{Pr}_{\text{CVS}}(\bar x \vert\bar { n } ) }  \right]  } 
\end{equation}
where $\bar x=(x_1,\ldots,x_{2M})=(q_1,\dots,q_M,p_1,\dots,p_M)$ is the continuously distributed measurement outcome.
This model of detection is equivalent to perfect eight-port homodyne detection, followed by a binning of the outcome results performed at the stage of post-processing. 
We assume a scaling of the window size $\eta$ with the number of modes, namely $\eta=2^{-poly(M)}$, analogous to what has been done in Ref.~\cite{Chakhmakhchyan2017}.
%The finitely resolved $ q$-homodyne operator as
%\be \label{operator-proj} {q}^{\eta} = \sum_{k = - \infty}^{\infty} q_k \int_{- \infty}^{\infty} \di q \chi^\eta_k(q) \vert q \rangle \langle q \vert \equiv  \sum_{k = - \infty}^{\infty} q_k {Q}_k \ee
%with $\chi^\eta_k(q) = 1$ for $p \in [q_k - \eta, p_k + \eta]$ and $0$ outside, $q_k = 2 \eta k$ and $2 \eta$ the resolution, associated with the width of the detector pixels. 

Our aim is to prove that the probability distribution $\text{Pr}^\eta_{\text{CVS}}(\bar { b }| \bar { m })$, where $\bar m=(1,\ldots,1,0,\ldots,0)$ describes the input photon subtractions, is hard to sample for a classical computer, both in the worst case scenario %(i.e. among the class of matrices $Q$, some of them will yield hard-to-sample output probability distribution) 
and in the average case scenario. %(i.e. extracting randomly $Q$, the output probability distribution is hard to sample with high probability).
In the next section we outline the proof. 

%-------------------------------------------------------------------------------------------------------------------------------------------------------------------------
\section{Auxiliary models and structure of the proof of hardness}
\label{se-proof}

%--------------------------------------------------------------------------------------------------------------
\begin{figure}
\begin{equation*}
\Qcircuit @C=1.2em @R=0.3em {
    \lstick{\ket{1}} & \multigate{5}{\phantom{aa}S(s)\phantom{al}} & \multigate{5}{\phantom{aa} {Q}\phantom{al}} & \measureD{} & \\
     \overset { \vdots  }{\phantom{a}}\phantom{blabi} & \ghost{\phantom{aa}S(s)\phantom{al}} & \ghost{\phantom{aa} {Q}\phantom{al}} &  \measureD{} & \rstick{\text{eight-port}}\\
    \lstick{\ket{1}} & \ghost{\phantom{aa}S(s)\phantom{al}} & \ghost{\phantom{aa} {Q}\phantom{al}}  &\measureD{} & \rstick{\text{homodyne}}\\
    \lstick{\ket{0}} & \ghost{\phantom{aa}S(s)\phantom{al}} & \ghost{\phantom{aa} {Q}\phantom{al}} &\measureD{} & \rstick{\text{detection}}\\
     \overset { \vdots  }{\phantom{a}}\phantom{blabi} & \ghost{\phantom{aa}S(s)\phantom{al}} & \ghost{\phantom{aa} {Q}\phantom{al}} & \measureD{} & \\
    \lstick{\ket{0}} & \ghost{\phantom{aa}S(s)\phantom{al}} & \ghost{\phantom{aa} {Q}\phantom{al}} & \measureD{} &
}
\end{equation*}
\caption{An alternative representation of a CVS circuit. The input has been rewritten using the mapping of single photon subtraction onto squeezing applied to single photons. $S(s)$ is the unitary associated to a squeezing with parameter $s$, while $Q$ is a passive linear optics transformation. The output is measured using eight-port homodyne detection.}
\label{fig:first-mapping}
\end{figure}
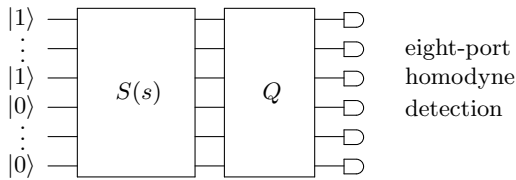

%--------------------------------------------------------------------------------------------------------------

In order to prove the hardness of the model that we are interested in, we define auxiliary computational models that we address as intermediate steps. First, we note that the input photon-subtracted squeezed states can be mapped onto single photon states, followed by squeezing. Indeed the squeezing operator $\hat{S}(s)$ satisfies $\hat{S}^{\dag}(s)\hat{a}\hat{S}(s)=c_s\hat{a}-s_s\hat{a}^\dag$ with $c_s=\cosh(\ln s)%=\frac{1}{2} (l+\frac{1}{l})
$ and $ s_s=\sinh(\ln s)%=\frac{1}{2} (l -\frac {1}{l})
$.
Therefore we obtain
\begin{eqnarray}
\label{eq:first-mapping}
\hat{a}\ket{s} =c_s\hat{S}(s)\hat{a}\ket{0}-s_s\hat{S}(s)\hat{a}^{\dag}\ket{0} = -s_s\hat{S}(s)\ket{1},
\end{eqnarray}
where the factor $s_s$ stems from the normalization of the left hand side of Eq.(\ref{eq:first-mapping}).
As a consequence, the circuit represented in Fig.\ref{fig:CVS} is fully equivalent to a circuit with input single photon states in correspondence with the photon-subtractions, followed by a squeezing operator $\hat{S}(s)$ applied to all the modes (Fig.\ref{fig:first-mapping}). By virtue of the identity in Eq.(\ref{eq:first-mapping}), the two architectures share the same probability distribution of the measurement outcomes $\text{Pr}^\eta_{\text{CVS}}(\bar { b }| \bar { m })$ in Eq.(\ref{distribution}), and therefore we refer to both configurations by CVS.

Next, we use time-reversal, i.e. the symmetry of Born's rule, in order to relate the probability distribution of CVS circuits to a matrix permanent. Because of the symmetry of Born's rule, the role of measurements and input states can be interchanged, while the probability distribution remains identical: input single photon states correspond to detection of single photons in the output modes and eight-port homodyne detection corresponds to input squeezed states, with a squeezing parameter $k \equiv 1/r$. The unitary evolution in between the input state and the output measurement is conjugated. Therefore, in the time-reversed version of the circuit in Fig.(\ref{fig:first-mapping}), squeezing occurs in the opposite quadrature, i.e. with parameter $l \equiv 1/s$, and is preceded by the passive linear optics evolution $T = Q^{\dagger}$.

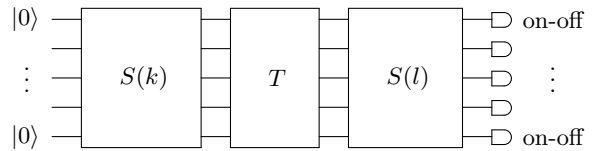
\begin{figure}
\begin{equation*}
\Qcircuit @C=1.2em @R=0.3em {
    \lstick{\ket{0}} & \multigate{4}{\phantom{aa}S(k)\phantom{al}} & \multigate{4}{\phantom{aa} T\phantom{al}} & \multigate{4}{\phantom{aa}S(l)\phantom{al}} & \measureD{} & \quad\text{on-off} \\
    & \ghost{\phantom{aa}S(k)\phantom{al}} & \ghost{\phantom{aa}T\phantom{al}} & \ghost{\phantom{aa}S(l)\phantom{al}} & \measureD{} & \\
     \overset { \vdots  }{\phantom{a}}\phantom{blabi} & \ghost{\phantom{aa}S(k)\phantom{al}} & \ghost{\phantom{aa} T\phantom{al}} & \ghost{\phantom{aa}S(l)\phantom{al}} & \measureD{} & \quad\overset { \vdots  }{\phantom{a}} \\
     & \ghost{\phantom{aa}S(k)\phantom{al}} & \ghost{\phantom{aa} T\phantom{al}} & \ghost{\phantom{aa}S(l)\phantom{al}} & \measureD{} & \\
    \lstick{\ket{0}} & \ghost{\phantom{aa}S(k)\phantom{al}} & \ghost{\phantom{aa} T\phantom{al}} & \ghost{\phantom{aa}S(l)\phantom{al}} & \measureD{} & \quad\text{on-off} \\
}
\end{equation*}
\caption{A representation of a TR-CVS circuit. The unitary matrix $T = Q^{\dagger}$ corresponds to a passive linear optics evolution, $Q$ as defined in Eq.(\ref{eq:definition-Q}), while $S(l)$ and $S(k)$ are the unitary matrices associated to the squeezing transformations with parameters $l = 1/s$ and $k=1/r$ respectively.}
\label{fig:TR-CVS_k}
\end{figure}

We define this model as ``Time-Reversed-Continuous-Variable Sampling" ($\text{TR-CVS}$). Its structure is outlined in Fig.~\ref{fig:TR-CVS_k}. 
By virtue of the symmetry of Born's rule,
\begin{equation}
\label{eq:equivalence-time-reversal}
\text{Pr}_{\text{CVS}}(0,\ldots,0|\bar{m})=\text{Pr}_{\text{TR-CVS}}(\bar{m})
\end{equation}
where $\text{Pr}_{\text{TR-CVS}}(\bar{m})$ is the probability of detecting the output pattern $\bar{m}$ in the TR-CVS circuit, and corresponds in the original circuit to the presence of $m$ single-photons (in turn related to single photon-subtractions in the original CVS model), while $\text{Pr}_{\text{CVS}}(0,\ldots,0|\bar{m})$ is the probability density function of the corresponding CVS circuit, evaluated at $(0,\ldots,0)$.
This allows us to establish a link between the output probability distribution of TR-CVS circuits and a matrix permanent. %, in analogy to what has been done in Ref.~\cite{Hamilton2016}.

%--------------------------------------------------------------------------------------------------------------
%--------------------------------------------------------------------------------------------------------------

%similar to GBS, but where the linear optics matrix evolution $U$ possesses a specific structure (Fig.\ref{fig:GBSlike}).
%More specifically, we show that for each evolution $T = Q^{\dagger}$ defined according to Eq.(\ref{eq:definition-Q}) and squeezing parameter $l$, even in the limit $k \rightarrow \infty$, we can always map TR-CVS circuits onto  GBS-like ones.

The proof of classical hardness for the probability distribution $\text{Pr}^\eta_{\text{CVS}}(\bar { b }| \bar { m })$ in Eq.(\ref{distribution}), then, is structured according to the following steps, that retrace the mappings outlined above:
\begin{enumerate}
\item We relate the probability distribution of TR-CVS circuits to a matrix hafnian. We then use time reversal symmetry Eq.(\ref{eq:equivalence-time-reversal}) to connect the probability distribution of TR-CVS circuits to the value $\text{Pr}_{\text{CVS}}(0,\ldots,0|\bar{m})$ of the probability density of CVS circuits. In addition to the properties of the matrix $\Sigma$ in Eq.(\ref{eq:definition-Q}), and to the relation between hafnian and permanent,  it yields that approximating multiplicatively $\text{Pr}_{\text{CVS}}(0,\ldots,0|\bar{m})$ is  \#P-hard (Theorem \ref{thExact}). 
%Using a conjecture on the permanent of real Gaussian matrices (Conjecture \ref{conj}),
\item We then show that sampling from $\text{Pr}^\eta_\text{CVS}(\bar { b }| \bar { m })$ allows for a  multiplicative approximation of $\text{Pr}_{\text{CVS}}(0,\ldots,0|\bar{m})$ in the third level of the polynomial hierarchy.  This will require making use of a Taylor expansion of the probability $\text{Pr}^\eta_\text{CVS}(\bar { b }| \bar { m })$, around $\bar { b }_0=(0,\ldots,0)$, for small $\eta$, as well as invoking the Stockmeyer's counting algorithm. Together with the previous point and standard complexity theory arguments -- see, e.g.~\cite{Aaronson2013,Bremner2015}, this allows us to claim the hardness of the original model in Fig.~\ref{fig:CVS}. The claim can then be made as an average statement (Theorem \ref{thMain}) using two conjectures.
\end{enumerate}

%-------------------------------------------------------------------------------------------------------------------------------------------------------------------------
\section{Proof of Hardness for CVS circuits}
\label{se:Hardness-proof}

We now explain in detail the proof of hardness sketched in the previous section. Let us first introduce our notations.

Any $M$-mode Gaussian state can be described by a $2M\times 2M$ covariance matrix $\sigma$ whose coefficients are defined for $k,l\in\{1,\ldots,2M\}$ by 
$\sigma_{kl}=\frac{1}{2}\left< \left\{ R_{ k }R_{ l } \right\}  \right> -\left< R_{ k } \right> \left< R_{ l } \right>$ where $\vec { R } =( { \hat{ q } } _{ 1 },\ldots, \hat{ q } _{ M }, \hat{ p } _{ 1 },\ldots, \hat{ p } _{ M })$ ~\cite{dutta1995real}. Alternatively and more conveniently, one can describe covariance matrices and evolutions in the complex basis  $\vec\Gamma={ ( \hat{ a }  }_{ 1 },\ldots,{  \hat{ a }  }_{ M }{ , \hat{ a }  }_{ 1 }^{ \dag  }{ ,\ldots, \hat{ a }  }_{ M }^{ \dag  })$, where ${  \hat{a}}_{j}=\frac {1}{\sqrt {2 }  } ({  \hat{ q }  }_{ j }+i{  \hat{ p }  }_{ j })$ and ${  \hat{ a }  }_{ j }^{ \dag  }=\frac { 1 }{ \sqrt { 2 }  } ({  \hat{ q }  }_{ j }-i{  \hat{ p }  }_{ j })$ for all $j\in \{1,\ldots,M\}$.
The evolution of the covariance matrix during a Gaussian evolution (excluding displacements) is then described by the complex symplectic transformation
\begin{equation}
\sigma \rightarrow S\sigma S^{ \dag },
\end{equation}
where a complex symplectic matrix $S$ satisfies 
\begin{equation}
S\Omega_0J\Omega_0^\dag S^{ \dag }=\Omega_0J\Omega_0^\dag
\end{equation}
with $J=\begin{pmatrix} 0_M & { {1}_M } \\ { -{1}_M } & 0_M \end{pmatrix}$ and $\Omega_0=\frac{1}{\sqrt{ 2 }}\begin{pmatrix} 1_M & { i{1}_M } \\ { {1}_M } & -i1_M \end{pmatrix}$.
We will use the notations
\begin{equation}
{ S }_{ \xi}\equiv\begin{pmatrix} D_{ c }(\xi) & D_{ s }(\xi) \\ D_{ s }(\xi) & D_{ c }(\xi) \end{pmatrix}  
\end{equation}
with $D_{ c }(\xi)=\text{Diag}(\cosh(\ln\xi_1),\ldots,\cosh(\ln\xi_M))$ and $D_{ s }(\xi)=\text{Diag}(\sinh(\ln \xi_1),\ldots,\sinh(\ln \xi_M))$ for the symplectic matrices that implement squeezing and 
\begin{equation}
S_U\equiv\begin{pmatrix} U & 0_{ M } \\ 0_{ M } & U^* \end{pmatrix}
\end{equation}
with $U\in \mathcal{U}(M)$ for the symplectic matrix associated with passive linear evolutions.

Finally, for any symmetric $2M\times2M$-matrix $A=(a_{kl})_{1\le k,l \le2M}$, its hafinian is defined in Ref. \cite{caianiello1953quantum} as
\begin{equation}
\text{Haf}(A)=\sum _{ \left\{ { i }_{ 1 },{ i }_{ 2 } \right\} ,\dots ,\left\{ { i }_{ 2M-1 },{ i }_{ 2M } \right\}  }^{  }{ { a }_{ { i }_{ 1 }{ i }_{ 2 } }\dots { a }_{ { i }_{ 2M-1 }{ i }_{ 2M } } } ,
\label{Hafnian}
\end{equation}
where the sum is over the perfect matchings of the set $\{1,\dots,2M\}$.\\

We can now state our first result.
\begin{theo}\label{thExact}
Approximating multiplicatively the output probability density value ${\rm Pr}_{\rm CVS}(0,\ldots,0|\bar{m})$ of CVS circuits is \#P-hard.
\end{theo}
{\bf Proof:} The proof will be rather technical, and will rely on a series of results that we develop below. 

(i) Firstly, we rely on the analytic expression of the output probability of a measurement outcome $\bar{n}$ for TR-CVS circuits. It can be expressed as a function of the output covariance matrix $\sigma_{\rm out}$. Namely we have: 
\be
\label{eq:probability-xxx}
{\rm Pr}_{\text{TR-CVS}}(\bar{n})=\frac{{\rm Haf}(A_S)}{\bar{n}!\sqrt{{\rm det}({ \sigma }_{\rm out }+1_{2M}/2)}},
\ee
where $A_S$ is a submatrix of 
\be\label{eq:A}
A=\begin{pmatrix} 0_{ M } & 1_{ M } \\ 1_{ M } & 0_{ M } \end{pmatrix}\left[ 1_{ 2M }-{ ({ \sigma }_{\rm out }+\frac { 1 }{ 2 } 1_{ 2M }) }^{ -1 } \right]
\ee
obtained by keeping the $j^{th}$ and $M+j^{th}$ rows and the $j^{th}$ and $M+j^{th}$ columns only if a photon has been detected in the $j^{th}$-mode. For the sake of clarity of the present manuscript, we refer to~\cite{Hamilton2016} for the detailed derivation of this expression.\\

(ii) Secondly, we focus on the output probability density of CVS circuits. Using Eq.(\ref{eq:equivalence-time-reversal}), this requires in turn deriving an expression for the matrix $A$ in Eq.\eqref{eq:probability-xxx}.
%\textcolor{red}{We are now interested in computing explicitly the output probability density of CVS circuits evaluated in $(0,...,0)$. Using Eq.(\ref{eq:equivalence-time-reversal}), this requires in turn deriving an expression for the matrix $A$ in Lemma~\ref{lemTRCVS}. We thereby aim at proving the following result:}
%We now seek for an explicit expression . {Note that in the case of TR-CVS circuits the covariance matrix  at the output $\sigma_{\rm out}$ is obtained after applying a squeezing transformation, followed by the unitary transformation $T$, followed by a second squeezing operation, departing from the structure in Ref.~\cite{Hamilton2016}.
%We will also link it back to CVS circuits using the following result:

For TR-CVS circuits, the symplectic matrix describing the evolution reads
\begin{equation}
\begin{aligned}
{S}_{\text{TR-CVS}}&=S_l S_T S_k \\
&=\begin{pmatrix} D_{ c }(l) & D_{ s }(l) \\ D_{ s }(l) & D_{ c }(l) \end{pmatrix}\begin{pmatrix} T & 0_{ M } \\ 0_{ M } & T^{ * } \end{pmatrix}\begin{pmatrix} D_{ c }(k) & D_{ s }(k) \\ D_{ s }(k) & D_{ c }(k) \end{pmatrix} \\
\end{aligned}
\label{EulerCVS}
\end{equation}
where $T$ is a $M\times M$ unitary matrix. The covariance matrix evolves according to
\begin{equation}
{\sigma }_{\rm out }= { S }_{\text{TR-CVS}}{ \sigma  }_{\rm  in } { S }_{\text{TR-CVS}}^{ \dag }.
\end{equation}
The input being a vacuum product state, ${ \sigma  }_{ in }={ \sigma  }_{ vacuum }=1_{ 2M } /2$,
we obtain 
\begin{equation}
\label{eq:cov-output}
\begin{aligned} 
{\sigma }_{\rm out }=\frac { 1 }{ 2 } { S }_{\text{TR-CVS}}{ S }_{\text{TR-CVS}}^{ \dag }.
\end{aligned}
\end{equation} 
From Eqs.\eqref{eq:A} and~\eqref{eq:cov-output} we obtain
\begin{equation}
A = B^*\oplus B,
\label{A}
\end{equation}
where 
\begin{equation}
\label{eq:boh2}
\begin{aligned}
B=&[D_{ c }(l)TD_{ s }(k)+D_{ s }(l)T^*D_{ c }(k)] \\
&\times[D_{ c }(l)T^* D_{ c }(k)+ D_{ s }(l)TD_{ s }(k)]^{-1}.
\end{aligned}
\end{equation}
Eq.(\ref{eq:boh2}) can be conveniently derived by using the Bloch-Messiah decomposition, which allows showing that the TR-CVS symplectic evolution in Eq.(\ref{EulerCVS}) is equivalent to squeezing followed by a unitary transformation. 
Let us introduce the notations $c_\chi=\cosh(\ln \chi)$, $s_\chi=\sinh(\ln \chi)$ and $t_\chi=\tanh(\ln \chi)$ for all $\chi\in\C$. With the equal squeezing assumptions, $D_{ c }(\chi)=c_\chi{1}_M$ and $D_{ s }(\chi)=s_\chi{1}_M$ for $\chi=k,l$, the matrix $B$ from the previous equation simply reads
\begin{equation}
\label{eq:boh3}
B=(c_ls_kT+s_lc_kT^*)(c_lc_kT^*+ s_ls_kT)^{-1}.
\end{equation}
The matrix $T$ describing the linear optics evolution of TR-CVS circuits and defined as the conjugate of the matrix in Eq.(\ref{eq:definition-Q}) can be rewritten considering that
%--------------------------------------------
\ba
T &=& Q^{\dagger} = e^{i\phi\Sigma} O  \nn  \label{eq:T-structure} \\
&=& \cos\phi O+i\sin\phi\Sigma O  
%&=& \frac { 1 }{ \sqrt { 1+x^4 }  } (1_M+ix^2 \Sigma )O ,
\ea
with $O = \Theta^{ T } \in \mathcal{O}(M)$, and $\Sigma\in \mathcal{O}_S(M)$, where we have used $\Sigma^2=1_M$.%, and $\Sigma\leftrightarrow -\Sigma, O\leftrightarrow -O$ depending on the sign of $\cos\phi$ and $\sin\phi$.

%--------------------------------------------

We can take advantage of the specific structure in Eq.(\ref{eq:T-structure}) of the matrices $T$ in order to obtain the following developed expression for Eq.(\ref{eq:boh3}): 
\begin{equation}
\begin{aligned}
%B=\frac{1}{2}\frac { s_{k^2l^2}+x^4s_{l^2/k^2} }{ { c }_{ kl }^{ 2 }+{ x }^{ 4 }{ c }_{ k/l }^{ 2 } } {1}_M+i\frac { x^2{ s }_{ { k }^{ 2 } } }{ { c }_{ kl }^{ 2 }+{ x }^{ 4 }{ c }_{ k/l }^{ 2 } } \Sigma.
B & =  \frac{\left(l^4-k^4\right) \sin ^2\phi+\left(k^4 l^4-1\right) \cos ^2 \phi}{\left(k^2+l^2\right)^2 \sin ^2\phi+\left(k^2 l^2+1\right)^2 \cos ^2\phi} {1}_M \\
 & + i \frac{\left(k^4-1\right) l^2 \sin 2 \phi }{\left(k^2+l^2\right)^2 \sin ^2\phi+\left(k^2 l^2+1\right)^2 \cos ^2\phi} \Sigma.
\end{aligned}
\end{equation}
Note that the hafnian of a matrix does not depend on its diagonal entries. Since the submatrix $A_S$ is obtained by removing rows and columns with the same indices, its diagonal entries are diagonal entries of the matrix $A$. Hence we may subtract any diagonal matrix to $A$ and still obtain the same output probability distribution. In particular subtracting $\frac{\left(l^4-k^4\right) \sin ^2\phi+\left(k^4 l^4-1\right) \cos ^2 \phi}{\left(k^2+l^2\right)^2 \sin ^2\phi+\left(k^2 l^2+1\right)^2 \cos ^2\phi} {1}_{2M}$ yields 
\begin{equation}
\begin{aligned}
\label{eq:prefactor}
A' = i f(k,l,\phi)  ( -\Sigma \oplus  \Sigma)
%\begin{pmatrix} -\Sigma & 0 \\ 0 & \Sigma \end{pmatrix}.
\end{aligned}
\end{equation}
with $f(k,l,\phi)$ given by:
\be\label{eqf}
f(k,l,\phi) = \frac{\left(k^4-1\right) l^2 \sin 2 \phi }{\left(k^2+l^2\right)^2 \sin ^2\phi+\left(k^2 l^2+1\right)^2 \cos ^2\phi}.
\ee
We restrict now the attention to the configurations $\bar{n} = \bar{m}$, i.e. single photon detections in the first $m$ modes. 
Using  Eq.(\ref{eq:probability-xxx}) and noting that $\bar{m}!= 1$, we finally obtain 
\ba
\text{Pr}_{\text{TR-CVS}}(\bar{m}) = \frac{f(k,l,\phi)^m }{\sqrt{\text{det}({ \sigma }_{\rm out }+1_{2M}/2)}} \text{Haf}(\Sigma_m)^2, 
%& & \text{Haf}(f(k,l,\phi) (-\Sigma_m \oplus \Sigma_m) ) \nn \\
 %\text{Haf}_m\begin{pmatrix} -\Sigma & 0 \\ 0 & \Sigma \end{pmatrix}  \nn \\
%&=&  \frac{f^2(k,l,\phi)}{\bar{m}!\sqrt{\text{det}({ \sigma }_{\rm out }+1_{2M}/2)}} \text{Haf}^2(\Sigma_m)
\label{Pr-TRCVS-final}
\ea
where we have also used that, for $m$ even, 
\begin{align}
\text{Haf}(i f(k,l,\phi)& (-\Sigma_m \oplus \Sigma_m) ) \nn \\
&=   \text{Haf}(-i f(k,l,\phi )\Sigma_m )\text{Haf}( i f(k,l,\phi) \Sigma_m) \nn \\
&=   f(k,l,\phi)^m \text{Haf}(\Sigma_m)^2.
\end{align}
 %Here Haf$_m$ indicates that we are interested in the submatrix obtained by keeping, for $j\in\{1,\ldots,m\}$, the $j^{th}$ and $M+j^{th}$ rows and $j^{th}$ and $M+j^{th}$ columns, and where with an abuse of notation 
$\Sigma_m$ indicates the $m \times m$ submatrix of $\Sigma$ at the upper left corner. %The symbol $\propto$ indicates that the lhs and rhs are equal up to an efficiently computable coefficient.

We can now use the time-reversal symmetry to obtain the expression for the output probability density of CVS circuits evaluated at $(0,...,0)$. Injecting Eq.(\ref{Pr-TRCVS-final}) in Eq.\eqref{eq:equivalence-time-reversal} we get
\be
\label{Pr-CVS-final}
{\rm Pr}_{\rm CVS}(0,\ldots,0|\bar{m}) = \frac{f(k,l,\phi)^m }{\sqrt{{\rm det}({ \sigma }_{\rm out }+1_{2M}/2)}}   {\rm Haf}(\Sigma_m)^2.
\ee
%
%\ba \hspace{-0.35cm} \text{Pr}_{\text{CVS}}(0,\ldots,0|\bar{m}) = \frac{f(k,l,\phi)^m }{\sqrt{\text{det}({ \sigma }_{\rm out }+1_{2M}/2)}}   \text{Haf}^2(\Sigma_m). \label{Pr-CVS-final} \ea
Thus we have shown that the output probability density for the specific input $\bar m$ is, at the origin, directly proportional to the hafnian squared of the upper left submatrix of a matrix describing the linear optical network. \\

(iii) Next, we want to relate the output probability of CVS circuits to the permanent of a real matrix. Specifically, we provide an explicit construction holding for any $m/2\times m/2$ real square matrix $X$: for all $M\geq 2m$ and $\nu\leq 1/\vert\vert X\vert\vert$ there exists a matrix $\Sigma\in\mathcal{O}_S(M)$ such that its top-left $m\times m$ submatrix is 
\begin{equation}\label{eq:Sigmam}
\Sigma_m=\nu\begin{pmatrix} 0 & X \\ X^{ T } & 0 \end{pmatrix}.
\end{equation}
In order to show this result, we define $Y=\nu X$. Observe that $1_{m/2}-Y^TY$ is symmetric positive semidefinite since $\left\| Y \right\| \le1$. Thus it has a Cholesky decomposition $1_{m/2}-Y^{ T }Y=Z^{ T }Z$ for some square matrix $Z$. The columns of the $m\times \frac{m}2$ matrix $W=\begin{pmatrix} Y \\ Z \end{pmatrix}$ are then forming an orthonormal family that can be completed into an orthonormal basis of $\mathbb{R}^{r}$, for any $r\ge m$. The matrix obtained with these columns is orthogonal by construction and reads
\begin{equation}
\begin{pmatrix} Y & C \\ B^{ T } & D \end{pmatrix}
\end{equation}
where $B, C$ are $m/2\times(r-m/2)$ matrices and $D$ is a $(r-m/2)\times(r-m/2)$ square matrix. Finally, by choosing $M\ge 2r$ and
\begin{equation}
\Sigma=\begin{pmatrix} 0 & Y & 0 & C & 0 \\ Y^{ T } & 0 & B & 0 & 0 \\ 0 & B^{ T } & 0 & D & 0 \\ C^{ T } & 0 & D^{ T } & 0 & 0 \\ 0 & 0 & 0 & 0 & 1_{M-2r}\end{pmatrix}
\end{equation}
we obtain an $M\times M$ symmetric orthogonal matrix (its columns are orthonormal by construction) which top left $m\times m$ submatrix is precisely given by Eq.(\ref{eq:Sigmam}),
%
%\begin{equation} \Sigma_{m}=\nu \begin{pmatrix} 0 & X \\ X^{ T } & 0 \end{pmatrix},\end{equation}
%
with the constraint $M\ge2m$. \\

(iv) Finally, recall that a specific relation holds between the hafnian and the permanent. For any square matrix $X$ we have:
\begin{equation}
\text{Perm}(X)=\text{Haf}\begin{pmatrix} 0 & X \\ X^{ T } & 0 \end{pmatrix}.
\label{permhaf}
\end{equation}
Using Eq.~\eqref{Pr-CVS-final} and the construction leading to Eq.\eqref{eq:Sigmam}, we get that for any square matrix $X$ there exists a CVS circuit which probability density at the origin reads:
\begin{equation}
\begin{aligned}
%\text{Pr}_{\text{CVS}}(0,\ldots,0|\bar{m})  
\text{Pr}_{\text{CVS}}(0,\ldots,0|\bar{m})  = \frac{f(k,l,\phi)^m \nu^{m} }{\sqrt{\text{det}({ \sigma }_{\rm out }+1_{2M}/2)}} \text{Perm}(X)^2.
\label{eq:probabilita-finale}
\end{aligned}
\end{equation}
It was shown in~\cite{Aaronson2013} that multiplicative approximation of ${\rm Perm}(X)^2$ is a \#P-hard problem for real square matrices. It implies that multiplicative approximation of ${\rm Pr}_{\rm CVS}(0,\ldots,0|\bar{m}) $ is also, provided the multiplying factor is finite and non-vanishing. The determinant can be derived using the same technique as for Eq.\eqref{eq:boh2}, yielding
{\small \be
\text{det}({ \sigma }_{\rm out }+\frac121_{2M}) =\left[\frac{\left(k^2+l^2\right)^2 \sin ^2(\phi )+\left(k^2 l^2+1\right)^2 \cos ^2(\phi )}{4 k^2 l^2}\right]^M.
\ee}
Using $f(k,l,\phi)$ given by Eq.\eqref{eqf}, we have overall an explicit expression for the coefficient that appears in Eq.~\eqref{eq:probabilita-finale} that only depends on $k,l$ and $\phi$. It is clear that it is finite and non-vanishing for some values of $k,l$ and $\phi$. This concludes the proof of Theorem~\ref{thExact}.\qed\\

We stress that the matrix $\Theta$ appearing in the definition of the CVS circuit Eq.(\ref{eq:definition-Q}) (and consequently the matrix $O$ in Eq.(\ref{eq:T-structure})) does not contribute to the output probability distribution Eq.(\ref{eq:probabilita-finale}). It provides additional degrees of freedom that can be particularly useful for experimental considerations -- see also the discussion in Section~\ref{se:Exp}.

%----------------
%Worst case
%
%\footnote{A problem $p$ is called hard for a complexity class $C$ if there exists a reduction from any problem in $C$ to $p$. The complexity class \#P is the class of problems that give the number of solutions of an NP problem.}. 
%
%"adaptive multiplicative estimates"  veut dire que l'algorithme ne te donne pas la valeur exacte du permanent avec un nombre polynomial d'estimations donne, mais plutot en faisant des estimations successives, chacune biais\UTF{00E9}e potentiellement par le resultat des precedented (d'ou le terme "adaptive"), en nombre polynomial on obtient la valeur exacte du permanent (tu as le detail de la construction p.32-34 de BosonSampling, theoreme 28)

%----------------
%Average case
This statement is a``worst-case" statement, in the sense that, as we have shown, for each matrix $X$ it is possible to find an instance of a CVS circuit $\mathcal{C}_X$ where the probability density evaluated at zero is a multiplicative approximate of $\text{Perm}(X)^2$. Hence this result states that it is the estimation of possibly all the output probabilities of the CVS circuits which is a $\#$P-hard problem.% In this sense, it is the approximation of all the values $\text{Pr}_{\text{CVS}}(0,\ldots,0|\bar{m})$ stemming from the full class of CVS circuits that defines a \#P-hard problem.

In order to strengthen our results and identify which fraction of CVS circuits are hard, we define the Real Gaussian Permanent Estimation problem:
\begin{pb}\label{pb}
{\rm(RGPE)} Given as input a matrix $X\sim \mathcal{N}(0,1)_\mathbb{R}^{p\times p}$ of i.i.d. Gaussians together with error bounds $\epsilon,\delta>0$, estimate ${\rm Perm}(X)$ to within error $\pm\epsilon\cdot |{\rm Perm}(X)|$, with probability at least $1-\delta$ over $X$, in ${\rm poly}(p, 1/\epsilon, 1/\delta)$ time.
\end{pb}
Recall that we can use the construction developed in (iii) of the proof of Theorem~\ref{thExact} for the particular case of i.i.d. Gaussian matrices: for any $X\sim \mathcal{N}(0,1)_\mathbb{R}^{p\times p}$ of i.i.d. Gaussians, we obtain a CVS circuit $\mathcal{C}_X$ such that Eq.~\eqref{eq:probabilita-finale} holds. Hence every instance of the RGPE is associated with a specific CVS circuit.
In relation to the problem above, we assume the \emph{Permanent-of-Real-Gaussians} Conjecture, or PRGC : 
\begin{conj}
\label{conj1}
{\rm RGPE} is \#P-hard.
\end{conj}

We will also need a second conjecture which is precisely the same \emph{Permanent Anti-Concentration} Conjecture as in~\cite{Aaronson2013} that we will apply to real matrices:
\begin{conj}
\label{conj2}
There exists a polynomial $P$ such that for all $n$ and $\delta >0$, 
\be
{\rm Pr}_{X\sim \mathcal{N}(0,1)_\mathbb{C}^{p\times p}}\left[\vert{\rm Perm}(X)\vert<\frac{\sqrt n!}{P(n,1/\delta)}\right]<\delta.
\ee
\end{conj}

This leads us to our second and more important result.
\begin{theo}\label{thMain}
%For all $X\sim \mathcal{N}(0,1)_\mathbb{R}^{p\times p}$ of i.i.d. Gaussians, assume Conjecture~\ref{conj2} is true, then we can use $\mathcal{C}_X$ to solve RGPE$_X$ in the third level of the Polynomial Hierarchy. Assume Conjecture~\ref{conj1} is true, then an efficient classical simulation of $\mathcal{C}_X$ would imply a collapse of the Polynomial Hierarchy.

\begin{enumerate}[label=(\alph*)]
\item \label{(a)} Assuming Conjecture~\ref{conj2} is true, CVS circuits can be used to solve RGPE in the third level of the Polynomial Hierarchy.
\item \label{(b)} Assuming Conjecture~\ref{conj1} is also true, then an efficient classical simulation of CVS circuits would imply a collapse of the Polynomial Hierarchy to its third level.
\end{enumerate}

\end{theo}
{\bf Proof:} \ref{(a)} As we have seen, the eight-port homodyne detections that enter the definition of CVS circuits are characterized by a finite resolution $\eta$. We will show that sampling from the probability distribution ${\rm Pr}_{\rm CVS}^\eta$ enables multiplicative approximation of the value of the probability density ${\rm Pr}_{\rm CVS}(0,\ldots,0|\bar{m})$ hence solving Problem~\ref{pb}.
%It was shown in~\cite{Aaronson2013} that for $p=O(n^{1/6})$, a $p\times p$ submatrix of a $n\times n$ Haar-random orthogonal matrix is close to a matrix of i.i.d. real Gaussian entries. It is also believed that $p=O(\sqrt n)$ would be sufficient, and in the following we will assume that the latter holds true.

%We pick Haar-randomly an $M\times M$ orthogonal matrix. Block-wise it reads:
%\begin{equation}
%O'=\begin{pmatrix} X & C \\ B^{ T } & D \end{pmatrix},
%\end{equation}
%where $X$ is an $m/2\times m/2$ matrix. Since we suppose in our model that $m=O(\sqrt M)$, $X$ is close to a matrix of i.i.d. real Gaussian entries. Using the same construction as in Lemma~\ref{lemPerm}, we obtain $\Sigma$ symmetric orthogonal with its top-left $m\times m$ submatrix being
%
%\begin{equation}
%\Sigma_m=\begin{pmatrix} 0 & X \\ X^{ T } & 0 \end{pmatrix}
%\end{equation}
%
%and for the corresponding circuits,
%
%\begin{equation}
%\label{eq:prob-density}
%\text{Pr}_{\text{CVS}}(0,\ldots,0|\bar{m}) \propto\text{Perm}^2(X).
%\end{equation}
%
%Hence for any Haar-random orthogonal matrix $O'$ one can construct a symmetric orthogonal matrix $\Sigma$ associated with a CVS circuit according to Eq.(\ref{eq:T-structure}). With the PRGC and Eq.(\ref{eq:prob-density}), this circuit results  with high probability in an expression for $\text{Pr}_{\text{CVS}}(0,\ldots,0|\bar{m})$ that is directly related to the RGPE. 

The probability distribution for finite resolution, given in Eq.(\ref{distribution}), reads or the input state described by $\bar { m }$
\begin{equation}
\text{Pr}^\eta_{\text{CVS}}(\bar { b }| \bar { m })\equiv \prod _{ i=1 }^{ M }{ \left[ \int _{ B^{(q)}_i  }\int _{B^{(p)}_i  }{ \mathrm d{ q }_{ i }\mathrm d{ p }_{ i }\text{Pr}_{\text{CVS}}(\bar x \vert\bar { m } ) }  \right]  } .
\end{equation}
As an intermediate step of the proof, we focus on the discrete outcome $\bar { b }_0 = (0,\ldots,0)$. Following the approach of Ref.~\cite{Chakhmakhchyan2017}, we perform a Taylor expansion of the multivariate function
\begin{equation}
f:\bar{x}=(x_1,\ldots,x_{2M}),\mapsto f(\bar x)=\text{Pr}_{\text{CVS}}(\bar{x}|\bar { m } )
\end{equation}
where $\bar{x}=(q_1,\ldots,q_M,p_1,\dots,p_M)$, around the value $\bar{x}_0=(0,\dots,0)$. This allows the finite resolution probability $\text{Pr}^\eta_{\text{CVS}}(\bar { b }_0| \bar { m })$ to be related to the expression $\text{Pr}_{\text{CVS}}(0,\ldots,0|\bar { m })$.
Assuming that $\eta$ is sufficiently small, we keep terms up to the second order in the series expansion of $\text{Pr}_{\text{CVS}}^\eta(\bar{b}|\bar { m } )$. After integration we get
\begin{equation}
\begin{aligned}
\text{Pr}^\eta_{\text{CVS}}(\bar { b }_0| \bar { m }) &= \eta^{2M}f(\bar{x}_0)\\+\frac{\eta^{2M+2}}{24}&\sum _{ i,j=1 }^{ 2M }{ \frac { \partial^2 f }{ \partial { x }_{ i } \partial x_j} (\bar{x}_0) } + O(\eta^{2M+4}).
%&=\eta^M\left[f(\bar{p}_0)+\frac{\eta^2}{24}\sum _{ i,j=1 }^{ M }{ \frac { \partial^2 f }{ \partial { p }_{ i } \partial p_j} (\bar{p}_0) } + O(\eta^4)\right],
\end{aligned}
\label{Taylor}
\end{equation}

Now we make use of Stockmeyer's approximate counting algorithm~\cite{stockmeyer1985approximation}. It is a classical algorithm that yields multiplicative estimates of a probability based on samples of the global probability distribution (a clear and compact explanation is provided e.g. in Ref.~\cite{Lund2017}). 
It %belongs to the complexity class $\text{FBPP}^\text{NP} $, where BPP is the class of probabilistic polynomial-time algorithms succeeding at least $2/3$ of the times, the prefix F indicates that the output is a function (a number) rather than a decision (yes/no), and the superscript notation describes access to an NP oracle. NP is the class of decision problems which solutions can be verified in polynomial time. Since $\text{BPP} \subseteq \text{NP}^\text{NP}$, Stockmeyer's algorithm 
is contained within the third level of the polynomial hierarchy~\cite{lautemann1983bpp}.

Having at our disposal an oracle that samples from the probability distribution $\text{Pr}^\eta_{\text{CVS}}(\bar { b }| \bar { m })$ allows for the approximation of the probability $\text{Pr}^\eta_{\text{CVS}}(\bar { b }_0| \bar { m })$ to within a multiplicative error by making use of Stockmeyer's algorithm, in the third level of the polynomial hierarchy. 
%
%{\color{red}Furthermore, choosing the resolution $\eta$ such that the first order term is negligible compared to the zeroth, i.e. for a given $\epsilon'>0$,
%\begin{equation}
%\label{eq:requirement-eta}
%\eta^2\le \frac{24f(\bar{x}_0)}{{ \sum _{ i,j=1 }^{ 2M }{ \frac { \partial^2 f }{ \partial { x }_{ i } \partial x_j} (\bar{x}_0) } }}\epsilon',
%\end{equation}
%
%we observe thanks to Eq. \eqref{Taylor} that the multiplicative-error estimate for $\text{Pr}^\eta_{\text{CVS}}(\bar { b }_0| \bar { m })$ provided by Stockmeyer's algorithm translates into a multiplicative-error estimate for $f(\bar{x}_0)=\text{Pr}_{\text{CVS}}(0,\ldots,0|\bar { m })$ without changing the level of the polynomial hierarchy which this algorithm is contained in. The scaling of $\eta=2^{-poly(M)}$ that we assumed in the definition of the model allows for the requirement in Eq.(\ref{eq:requirement-eta}) to be fulfilled. A careful analysis of the summation in Eq.(\ref{eq:requirement-eta}) could yield to a less stringent requirement on the needed scaling, but we leave this analysis for future work.} \\
The resolution $\eta$ must be small enough to ensure that higher order terms are negligible compared to the term of order zero in the Taylor expansion. However, Stockmeyer's algorithm fails if the probability it is supposed to estimate is too small. Thus, $\eta$ must also be large enough to ensure that Stockmeyer's algorithm does not fail. Using Conjecture~\ref{conj2}, it is shown in Ref.~\cite{Lund2017} that a scaling $\eta=2^{-poly(M)}$ is fulfilling both requirements. While the Taylor expansion is slightly different in our case, the same argument holds, and assuming $\eta=2^{-poly(M)}$ implies that Stockmeyer's algorithm leads to a multiplicative error estimate for $f(\bar{x}_0)=\text{Pr}_{\text{CVS}}(0,\ldots,0|\bar { m })$ with an algorithm still in the third level of the polynomial hierarchy.
%Haence assuming this scaling of $\eta$, the multiplicative-error estimate for $\text{Pr}^\eta_{\text{CVS}}(\bar { b }_0| \bar { m })$ provided by Stockmeyer's algorithm translates into a multiplicative-error estimate for $f(\bar{x}_0)=\text{Pr}_{\text{CVS}}(0,\ldots,0|\bar { m })$ without changing the level of the polynomial hierarchy which this algorithm is contained in. }

We now have all the ingredients to conclude on the classical hardness of our model. 
We follow the same reasoning as in Ref.~\cite{Aaronson2013, Chakhmakhchyan2017}.
%\begin{theo}\label{thMain3}
%Suppose there exists an efficient classical algorithm able to approximate multiplicatively the output distribution of {\rm CVS} circuits. Then the Polynomial Hierarchy collapses to the third level.
%Here
%\end{theo}
Assume there exists an oracle $\mathcal{O}$ which, given the description of the CVS circuit and a random string $r$ (as its only source of randomness), outputs a sample $\bar{b}$ according to the distribution $\text{Pr}^\eta_{\text{CVS}}(\bar { b }| \bar { m })$. 
%The probability that $\mathcal{O}$ outputs $\bar{b}_0=(0,\ldots,0)$ is then given as
%\begin{equation}
%\text{Pr}[\mathcal{O}(Q,l,r) = {\bar{b}_0}]=\text{Pr}^\eta_{\text{CVS}}(\bar { b }_0| \bar { m }).
%\end{equation}
For the above arguments, this would allow one to approximate $\text{Pr}_{\text{CVS}}(0,\ldots,0|\bar { m })$ to within a multiplicative error by means of Stockmeyer algorithm, i.e. in the third level of the polynomial hierarchy.

Let $X\sim \mathcal{N}(0,1)_\mathbb{R}^{\frac{m}2\times \frac{m}2}$ a square matrix which entries are i.i.d. Gaussians.  We saw while proving Theorem~\ref{thExact} that we could construct a matrix $\Sigma_m$, and thus a linear optical network for a CVS circuits, such that ${\rm Pr_{CVS}}(0,\ldots,0\vert\bar m)$ is proportional to ${\rm Perm}(X)^2$. Then for any matrix $X$ satisfying the hypotheses of Problem~\ref{pb} we may design a CVS circuit $\mathcal{C}_X$ to use as an oracle, and a classical algorithm such that we could approximate ${\rm Perm}(X)^2$ to within multiplicative error in the third level of the polynomial hierarchy.

Problem~\ref{pb} however refers to estimating the permanent of i.i.d. real Gaussians itself. It is easy to see that a multiplicative approximation of the permanent squared can be turned into a multiplicative approximation of the modulus of the permanent. Then in the case of real matrices only the sign of the permanent remains to be determined. A more general version of this question was already addressed in~\cite{Aaronson2013}  where they gave an elaborate reduction from multiplicative approximation of the permanent to additive approximation of the permanent squared based on Conjecture~\ref{conj2}. To do so they showed that Conjecture~\ref{conj2} allowed one to estimate the phase of ${\rm Perm}(X)$ from multiplicative approximation of $\vert{\rm Perm}(X)|^2$, for $X$ i.i.d. \emph{complex} Gaussian matrix. It means in particular that Conjecture~\ref{conj2} also allows one to determine the sign of ${\rm Perm}(X)$ from ${\rm Perm}(X)^2$ if $X$ is i.i.d. \emph{real} Gaussian matrix. So assuming Conjecture~\ref{conj2}, RGPE can actually be solved in the third level of the polynomial hierarchy using a cleverly designed CVS circuit as an oracle.

\ref{(b)}  Assuming Conjecture~\ref{conj1} is true, Problem~\ref{pb} is \#P-hard. The existence of an efficient classical algorithm to approximate multiplicatively the output distribution of {\rm CVS} circuits would immediately imply the existence of a classical algorithm sitting in the third level of the polynomial hierarchy able to solve a \#P-hard problem. This in turn would yield a collapse of the polynomial hierarchy to the third level, thanks to Toda's theorem \cite{Toda91}. This concludes the average-case hardness proof.
\qed\\

Given that the collapse of the polynomial hierarchy is considered highly unlikely, this provides a strong evidence for the classical hardness of simulatability  of CVS circuits. Note that worst-case hardness follows directly from Theorem~\ref{thExact} using the Taylor expansion and Stockmeyer's algorithm arguments developed in this section, without the use of the two conjectures.

\section{Consequences and extensions}
\label{se:remarks} 

In this section we list some relevant consequences of the results previously presented, and we consider some possible extensions as well as limiting cases of the sampling model.
The discussion will be based on the expression of the pre-factor appearing in Eq.(\ref{eq:prefactor}) and following ones, that relate the value of the probability density $\text{Pr}_{\text{CVS}}(0,\ldots,0|\bar{m}) $ to the permanent square of the submatrix $X$. Let us consider $\phi = \pi/4$ for simplicity. %(see Appendix \ref{app:phi} for the role of this parameter in the general case)
For this value we obtain 
\begin{align}
\label{eq:prefactor-simplified}
\kappa(k,l) &= \frac{f(k,l,\fr{\pi}{4})^m }{\sqrt{\text{det}({ \sigma }_{\rm out }+1_{2M}/2)}} \nn \\
&=\fr{  2^{m+\frac{3 M}{2}}\left(k^4-1\right)^m l^{2 m} \left( k l\right) ^M }{\left(k^4l^4+k^4+l^4+4k^2l^2+1\right)^{m+\frac{M}{2}}}.
\end{align}
Remark that Eq.(\ref{eq:prefactor-simplified}) is symmetric in $l \rightarrow 1/l$ and $k \rightarrow 1/k$, which physically corresponds to taking the opposite squeezing quadrature either at the input or for the output measurements, respectively.

%-------------------------------------------------------------------------------------------------------------------------------------------------------------------------
\subsection{Photon-added squeezed states}

\begin{figure}
\begin{equation*}
\Qcircuit @C=1.2em @R=0.3em {
    \lstick{{\hat a}^{\dagger}\ket{s}} & \multigate{5}{\phantom{aa}Q\phantom{al}} &\measureD{} &\\
     \overset { \vdots  }{\phantom{a}}\phantom{blabi} & \ghost{\phantom{aa}Q\phantom{al}} & \measureD{} & \rstick{\text{eight-port}}\\
    \lstick{{\hat a}^{\dagger}\ket{s}} & \ghost{\phantom{aa}Q\phantom{al}} &\measureD{} &\rstick{\text{homodyne}}\\
    \lstick{\ket{s}} & \ghost{\phantom{aa}Q\phantom{al}} &\measureD{} & \rstick{\text{detection}}\\
     \overset { \vdots  }{\phantom{a}}\phantom{blabi} & \ghost{\phantom{aa} Q \phantom{al}} & \measureD{} & \\
    \lstick{\ket{s}} & \ghost{\phantom{aa}Q\phantom{al}} & \measureD{} & 
}
\end{equation*}
\caption{Alternative family of CVS circuits, where single photon-subtraction on input squeezed states is replaced by single photon-addition.}
\label{fig:CVSbis}
\end{figure}
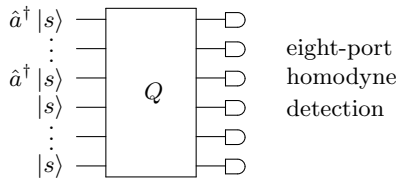

Input photon-subtracted squeezed states can be replaced by photon-added squeezed states, maintaining the non-Gaussian character of the input state, and resulting in the circuit of Fig.\ref{fig:CVSbis}. 
The full hardness proof that we have established in the previous sections goes through even in this case. Indeed one would obtain analogously to before $\hat{S}^{\dag}(s)\hat{a}^{\dag}\hat{S}(s)=-s_s\hat{a}+c_s\hat{a}^\dag$, i.e.
\begin{eqnarray}
\label{eq:first-mapping-2}
\hspace{-0.3cm} \hat{a}^{\dag}\ket{s} =-s_s\hat{S}(s)\hat{a}\ket{0}+c_s\hat{S}(s) \hat{a}^{\dag}\ket{0} = c_s\hat{S}(s)\ket{1} \hspace{-0.1cm},
\end{eqnarray}
which shows that the same mapping onto single photon states followed by squeezing can be performed, yielding back the circuit of Fig.\ref{fig:first-mapping}. 

%-------------------------------------------------------------------------------------------------------------------------------------------------------------------------
\subsection{Role of the squeezing parameter $s =1/l$ and limit of zero squeezing: Boson Sampling with eight-port homodyne detection}
\label{se:limit-squeezing}

We want here to discuss the role of the input squeezing parameter $s =1/l$. 
In the proof that we have presented, the only thing that depends on the squeezing degree $l$ is the prefactor in Eq.(\ref{eq:prefactor-simplified}).
As long as the parameter $l$ is constant with respect to the number of modes, like in the definition of our model, this pre-factor does not play any role for the argument yielding the hardness of CVS circuits. In particular, due to symmetry of Eq.(\ref{eq:prefactor-simplified}), the hardness result still holds considering opposite input squeezing quadratures, i.e. changing  $l \rightarrow 1/l$ in the equations. 

Our arguments also hold in the special configuration of zero squeezing $l =1$. Taking this limit in Eq.(\ref{eq:first-mapping}) we obtain
\be
-\frac{1}{s_l}\hat{a}\ket{l}\underset{l\rightarrow1}{\longrightarrow}\ket{1}
\ee
where the state on the rhs is the single photon Fock state $n = 1$. In accordance to our previous notation, the ket on the lhs $\ket{l}$ is a squeezed state whose squeezing parameter goes to $1$. 
%and this corresponds to the case of input single-photon states. 
Therefore we obtain, as a limiting case of our model, a hardness result for Boson Sampling with eight-port homodyne detection, and with unitary evolutions specified by Eq.(\ref{eq:definition-Q}).
Analogous models with input single-photon states have been demonstrated to be hard to sample: in Ref.~\cite{Chakhmakhchyan2017} with a distinct subclass of unitary matrices, and in Ref.~\cite{Lund2017}, with so-called CV-n measurements, that involve mixing the output modes with a $\ket{n}$ Fock state prior detection. 
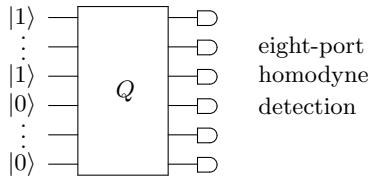
\begin{figure}
\begin{equation*}
\Qcircuit @C=1.2em @R=0.3em {
    \lstick{\ket{1}} & \multigate{5}{\phantom{aa} Q \phantom{al}} &\measureD{} & \\
     \overset { \vdots  }{\phantom{a}}\phantom{blabi} & \ghost{\phantom{aa} Q \phantom{al}} & \measureD{} & \rstick{\text{eight-port}}\\
    \lstick{\ket{1}} & \ghost{\phantom{aa} Q \phantom{al}} &\measureD{} & \rstick{\text{homodyne}}\\
    \lstick{\ket{0}} & \ghost{\phantom{aa} Q \phantom{al}} &\measureD{} & \rstick{\text{detection}}\\
     \overset { \vdots  }{\phantom{a}}\phantom{blabi} & \ghost{\phantom{aa} Q \phantom{al}} & \measureD{} & \\
    \lstick{\ket{0}} & \ghost{\phantom{aa} Q \phantom{al}} & \measureD{} & 
}
\end{equation*}
\caption{In the limit of zero squeezing, our model in Fig.\ref{fig:CVS} yields Boson Sampling with eight-port homodyne detection and unitary evolutions specified by Eq.(\ref{eq:definition-Q}).}
\label{fig:CVS-limit}
\end{figure}

%However, the value of $l$ affects the efficiency of the quantum sampler.
%As can be seen from Eq.(\ref{eq:prefactor}), the higher the value of the prefactor, the higher the sampling efficiency.
%The latter is maximized by the choice $l=1$, that corresponds to the case of zero input squeezing. The input squeezing therefore degrades the performance of the quantum sampler, with respect to the case of Boson Sampling with heterodyne detection.
As can be seen in Eq.(\ref{eq:prefactor}), the prefactor is maximized by the choice $l=1$, that corresponds to the case of zero input squeezing. The input squeezing therefore %degrades the performance of the quantum sampler
reduces the value of $\kappa(k,l)$, with respect to the case of Boson Sampling with eight-port homodyne detection.

%-------------------------------------------------------------------------------------------------------------------------------------------------------------------------
\subsection{Detection quadrature and projection over finitely squeezed states}

In a similar way to the preceding discussion, we may study the impact of the parameter $k$ on our proof. Recall that $k$ corresponds to the amount of squeezing of the basis states of the eight-port homodyne detection.  
%$k \rightarrow  \infty$ represents projection onto perfectly $ p$-squeezed states, while $k =  0$ means  that we exactly measure the quadrature  $ q$ by projection onto infinitely  $ q$-squeezed states. 

First, we can see that for $k=1$, i.e. when projecting onto coherent states via balanced eight-port homodyne detection, the pre-factor $\kappa(1,l)=0$. This implies that our proof does not hold anymore, and in general we cannot infer the hardness of the CVS circuits when based upon balanced eight-port homodyne detection.
Similarly, we cannot infer the hardness of our model for standard homodyne detection, that corresponds to a projection onto infinitely squeezed states and is parameterized by $k  \rightarrow \infty$ or $k  = 0$.

Nevertheless $\kappa(k,l)$ is non-vanishing for every finite $k>1$ (or $0 < k<1$, since $\kappa$ is invariant under $k\rightarrow1/k$). So our conclusion regarding the hardness of the CVS model remains valid for every unbalanced eight-port homodyne detection. 

For completeness, we study the maximum value of $\kappa$ seen as a function of $k$. In general, the maximum itself is a function of $l$.
Taking for simplicity $l=1$, the maximum of Eq.(\ref{eq:prefactor-simplified}) for $k>1$ is obtained for 
\be
k_0= \sqrt{1+\fr{2m+2\sqrt{m(m+M)}}{M}}.
\ee
Thus we showed that $\kappa(k,l)$  is optimal for a very specific unbalancing of the eight-port homodyne detection. 

%-------------------------------------------------------------------------------------------------------------------------------------------------------------------------
\section{Link with experiments}\label{se:Exp}

%The experiment that more closely relates to the model studied in the present paper, and that indeed was a motivation for developing it, is the one studied in Laboratoire Kastler Brossel~\cite{Ra2017, Roslund2013, Roslund_13b}. 
We would like to discuss here briefly the experiments~\cite{Ra2017, Roslund2013, Roslund_13b}, which are relevant to the model studied in the present paper. In that experimental configuration, several squeezed states are simultaneously available in the same optical cavity. Photon subtraction can be implemented mode-selectively, and detection can be performed simultaneously on all the optical models by means of multi-pixel homodyne detection~\cite{Beck00, ferrini2013compact}. The class of unitary transformations that can be implemented in the current version of the experiment has been characterized in Ref.~\cite{ferrini2013compact}, and amounts to unitary matrices of the form
\be
\label{eq:u-exp}
U_{\text{exp}} = O_{\text{post}} \Delta_{\text{LO}} O_{\text{change}} \Delta_{\text{OPO}}
\ee
with $O_{\text{post}}$ and $O_{\text{change}}$ real orthogonal matrices, and $\Delta_{\text{OPO}}$ and $ \Delta_{\text{LO}}$ diagonal matrices with modulo $1$ complex elements. The first two matrices $O_{\text{post}}$ and $ \Delta_{\text{LO}}$ are arbitrarily tunable, while in the current version of the experiment $\Delta_{\text{OPO}} = \text{diag}(1, -1, \ldots, 1, -1)$ and $O_{\text{change}}$ is also fixed. Physically, the matrix $\Delta_{\text{OPO}}$ means that the input squeezed states are squeezed on alternating quadratures. $O_{\text{change}}$ is implemented by detecting the optical modes in a mode basis that is different from the one in which the modes are individually squeezed.  

In order to establish a connection with this experiment, we rewrite the linear optics matrix of our model Eq.(\ref{eq:T-structure}) in yet another form, that renders transparent the comparison with Eq.(\ref{eq:u-exp}).
Since any real symmetric matrix can be orthogonally diagonalized thanks to the spectral theorem, and since the eigenvalues of a symmetric orthogonal matrix are either $1$ or $-1$, the matrix $\Sigma$ in Eq.(\ref{eq:T-structure}) can be written as
\begin{equation}
\Sigma=\omega\Delta\omega^{ T }
\label{diagsym}
\end{equation}
with $\omega\in \mathcal{O}(M)$ and $\Delta$ a diagonal matrix of $\pm1$'s. Let $p$ be the multiplicity of the eigenvalue $1$ of $\Sigma$ and $P$ the permutation matrix such that $P^{ T }\Delta P=\begin{pmatrix} {1}_{ p } & { 0 } \\ { 0 } & -{1}_{ M-p } \end{pmatrix}$. We have $PP^{ T }=1_M$ and then Eq. \eqref{eq:T-structure} reads
\begin{equation}
\begin{aligned}
\label{eq:confronto-exp}
T&=\cos\phi\omega PP^{ T }\Omega^{ T }O+i\sin\phi\omega\Delta\omega^{ T } O \\
&=\omega(\cos\phi PP^{ T }+i\sin\phi\Delta)\omega^{ T } O \\
&=O_1\begin{pmatrix} { e }^{ i\phi  }1_{ p } & { 0 } \\ { 0 } & { e }^{ -i\phi  }1_{ M-p } \end{pmatrix}O_2
\end{aligned}
\end{equation}
where $O_1=\omega P\in \mathcal{O}(M)$ and $O_2=P^{ T }\omega^{ T } O\in \mathcal{O}(M)$. 
Eq.(\ref{eq:confronto-exp}) explicits the gap (in terms of degrees of freedom) between the full unitary group and the class of matrices defined in our model. Indeed, any special unitary matrix can be decomposed according to the so-called $KAK$ decomposition~\cite{Helgason1978} as $U = O_1 \delta O_2$, with $\delta$ a general diagonal matrix of unit modulo complex elements.
 
Comparing Eqs.(\ref{eq:u-exp}) and (\ref{eq:confronto-exp}), we can readily see that if the matrix $\Delta_{\text{OPO}}$ was the identity, then the structure of experimentally implementable matrices would match the one defined in our model. Indeed, one could use the experimentally tunable degrees of freedom $O_{\text{post}} \Delta_{\text{LO}} $ to achieve any chosen $O_1\begin{pmatrix} { e }^{ i\phi  }1_{ p } & { 0 } \\ { 0 } & { e }^{ -i\phi  }1_{ M-p } \end{pmatrix}$ (which is fixed by the choice of $\Sigma$), and then adjust the matrix $O_2$ to recover $O_{\text{change}}$, yielding equivalence between Eq.(\ref{eq:u-exp}) and Eq.(\ref{eq:confronto-exp}).

The presence of alternating squeezing quadratures modeled by $\Delta_{\text{OPO}}$ forbids a straightforward application of our model in these experiments, and renders desirable an extension of our proof of hardness to the case of arbitrary input squeezing quadratures on each single mode. 
%Although not apparent from Eq.(\ref{eq:u-exp}), the squeezing levels in the experiment are not uniform over all the modes. However, upgrades of this experiment are being planned, where, applying pulse shaping techniques to the laser source that generates the squeezed states, the  squeezing spectrum could be partly tuned and thus made reasonably flat in the squeezing quadrature for the first, say, 50 modes
Although in actual experiments so far squeezing levels are not uniform, there are techniques for stabilizing them so that they can be made reasonably flat in the first, say, 50 modes~\cite{Arzani2017}.  
We expect therefore that (upon generalization to arbitrary input squeezing quadrature in each mode and homodyne detection) proofs of principle of our model could soon be implemented in that experiment. 

%This evolution corresponds to symplectic unitary matrices of the form
% \ba \hspace{-0.5cm}
% & & S_{linear}  = \\
%& & \begin{pmatrix} O_{ 1 }\begin{pmatrix} { e }^{ i\phi  }{1}_{ p } & 0 \\ 0 & { { e }^{ -i\phi  }{1}_{ M-p } } \end{pmatrix}O_{ 2 } & 0_{ M } \\ 0_M & O_{ 1 }\begin{pmatrix} { e }^{ -i\phi  }{1}_{ p } & 0 \\ 0 & { e }^{ i\phi  }{1}_{ M-p } \end{pmatrix}O_{ 2 } \end{pmatrix},  \nn  \\
%& &\begin{cases} \phi \in [0,2\pi ] \\ O_{ 1 },O_{ 2 }\in { O }(M) \\ p\in \{ 1,\ldots,M\}  \end{cases}. \nn  \label{class} \ea

%-------------------------------------------------------------------------------------------------------------------------------------------------------------------------
\section{Conclusions and perspectives}
\label{se:conclusions}

We have proven the computational hardness of a sampling problem that stems from a family of CV quantum circuits composed of photon-subtracted or photon-added squeezed states, linear optics evolution, and eight-port homodyne detection. 
Mapping to other sub-universal architectures, we could establish a worst-case proof that does not require any other conjecture than the fact that the polynomial hierarchy does not collapse. The average-case proof has required introducing two additional conjectures: the real version of the Permanent of Gaussian Conjecture and the Permanent Anti-Concentration Conjecture, both already present in standard Boson Sampling.

The main motivation to study this model comes from recent experimental results~\cite{Ra2017, Roslund2013}, which extensions could allow a proof of principle of the model outlined in this paper in the short-term. Furthermore, it is an interesting question to explore whether the use of squeezed states can bring some advantage with respect to the use of single photons. From the computational point of view, we have shown that this is not the case: squeezing is even unnecessary for classical hardness, in the sense that, as we have shown, the zero squeezing limit of our model reduces to Boson Sampling with eight-port homodyne detection, and such a model is still hard to sample.%However, the use of squeezed states offers a natural tool to increase the number of available input photons with respect to standard Boson Sampling. 
However, it would be interesting to investigate whether monitoring this squeezing could be used in order to lower the requirements to reach a regime where classical simulation cannot be achieved~\cite{Rahimi-Keshari2016}, that is nowadays still prohibitive in standard Boson Sampling experiments~\cite{Neville2017, Clifford2017}.

An important extension of the result presented in this work, that is especially necessary in view of the connection to experiments, is to provide a hardness proof for  sampling from an approximate probability distribution. We leave this extension to future work. 
Furthermore, we expect that the hardness of the scattershot version of our model~\cite{Lund2014}, corresponding to choosing randomly at each run of the sampling how the $m$ photon-subtractions are distributed among the $M$ optical modes, can also be proven. 

Finally, given the plethora of architectures which hardness has been demonstrated already, it becomes crucial to understand the general conditions required for a quantum advantage. Does the hardness result presented in this work still hold if we replace the input single-photon-subtracted states with other non-Gaussian states? % such as cat states, or even cubic phase states $e^{i  q^3 \theta} \ket{l}$ (with $\theta$ a real parameter)? 
 In the spirit of the discussion formulated at the end of Ref.~\cite{Seshadreesan2015}, we believe that it is fundamental to formulate a general sufficient condition. %, e.g. based on the negativity of the input Wigner function.
 %Olson2015: PASSV, Boson Sampling with photon-added and photon-subtracted 
%Seshadreesan2015: quello sulla transizione di fase (sempre displaced single photon )
%Rohde2015: cat states

%-------------------------------------------------------------------------------------------------------------------------------------------------------------------------
\section{Acknowledgements}

We kindly acknowledge L. Chakhmakhchyan and N. Cerf for having shared with us Ref.~\cite{Chakhmakhchyan2017} prior to its submission as well as for interesting discussions, and F. Arzani, V. Parigi, N. Treps, T. Ralph and S. Rahimi-Keshari for useful discussions.
This work was supported by the ANR COMB project, grant ANR-13-BS04-0014 of the French Agence Nationale de la Recherche, and by the DAAD-Campus France project Procope N$^\circ$ 35465RJ. G. F. acknowledges support from the European Union through the Marie Sklodowska-Curie grant agreement No 704192. 

%------------------------------------------------------------------------------------------------------------------------------------------------------------------------
\appendix 

\section{eight-port homodyne detection}
\label{app:heterodyne}

\begin{figure}
\includegraphics[width=0.8\columnwidth]{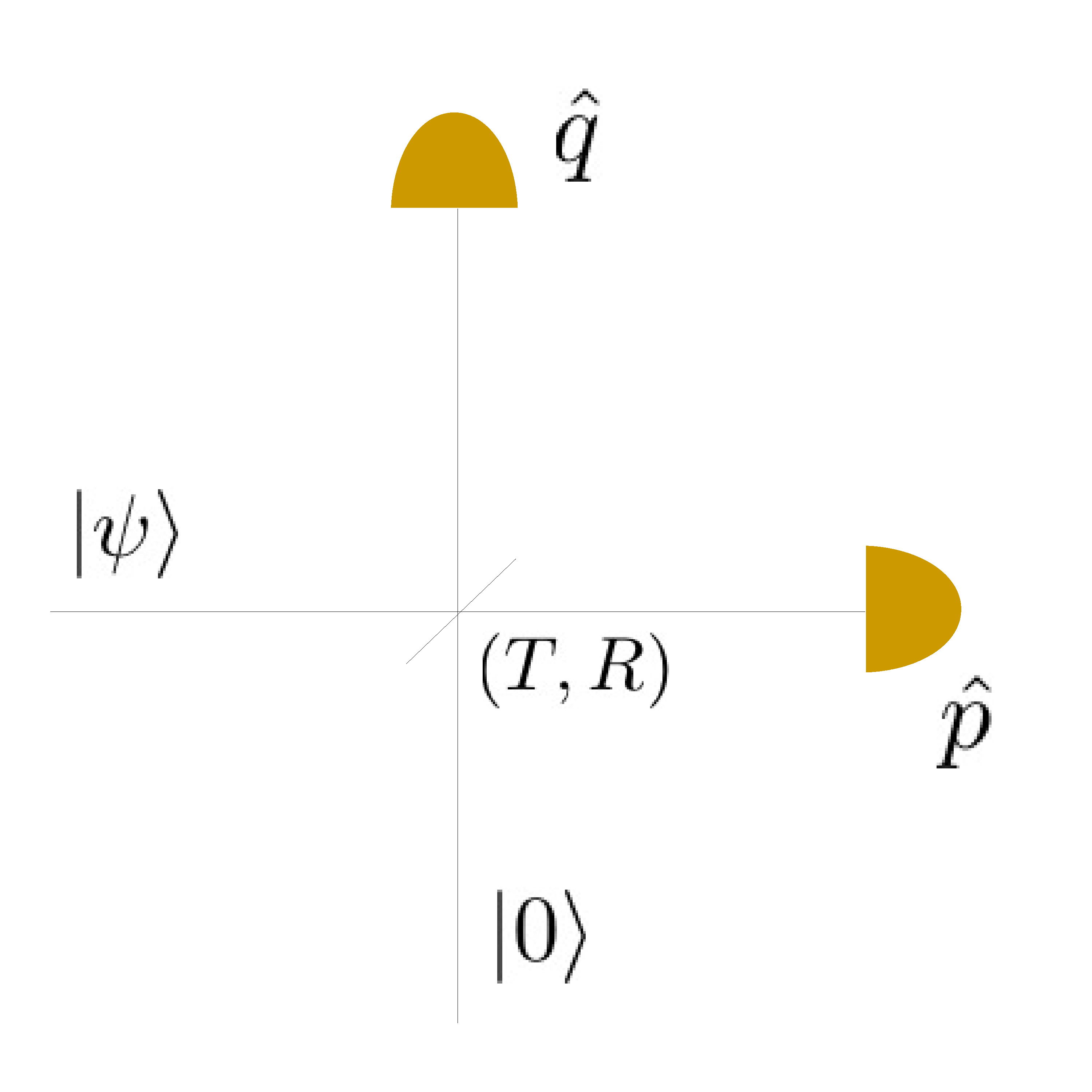}
\caption{Schematic representation of eight-port homodyne measurement. The $ \hat p$ and  $\hat q$ measurement are each performed by standard homodyne detection.}
\label{fig:heterodyne}
\end{figure}

In this appendix we present the modelization of the eight-port homodyne detection. The input state is mixed with vacuum at a beam splitter characterized by reflectivity $R$ and transmittivity $T$, with $R^2+T^2=1$.
We are interested in showing that this procedure results in projecting the state onto finitely squeezed states.
More precisely, we compute the POVM elements
\be
\Pi = \fr{1}{c} \ket{\psi(q_1,p_2)} \bra{\psi(q_1,p_2)}
\ee
with
\be
\label{eq:povm}
%\ket{\psi(q_1,p_2)} =\fr{1}{N} _2\bra{0} U_{\text{BS}} e^{-i  p_1 q_1} e^{i  q_2 p_2} \ket{q_1 =0,p_2=0},
\ket{\psi(q_1,p_2)} =\fr{1}{N} \bra{0} U_{\text{BS}} \ket{q_1,p_2},
\ee
where $\ket{q_1}$ and $\ket{p_2}$ are the position and momentum eigenstates
\begin{align}
\hat{q}_1 \ket{q_1} = \hat{q}_1 \ket{q_1}  \\
\hat{p}_2 \ket{p_2} = \hat{p}_2 \ket{p_2} 
\end{align}
and the symplectic action of a beam splitter is given by
\begin{align}
& U_{\text{BS}}  \begin{pmatrix}  \hat q_1   \\  \hat q_2\end{pmatrix}   U_{\text{BS}}^{\dagger} = \begin{pmatrix} T & -R \\ R &T \end{pmatrix}  \begin{pmatrix}  \hat q_1   \\  \hat q_2\end{pmatrix}  \\
& U_{\text{BS}}  \begin{pmatrix}  \hat p_1   \\  \hat p_2\end{pmatrix}   U_{\text{BS}}^{\dagger} = \begin{pmatrix} T & -R \\ R &T \end{pmatrix}  \begin{pmatrix}  \hat p_1   \\  \hat p_2\end{pmatrix}. 
\end{align}
$N$ is a normalization constant. 
Eq.(\ref{eq:povm}) can be directly computed by Fourier transforming the ket $\ket{p_2}= \int d q_2 e^{i q_2 p_2} \ket{q_2}$, yielding
\begin{align}
%\ket{\psi(q_1,p_2)} =\fr{1}{N} _2\bra{0} U_{\text{BS}} e^{-i  p_1 q_1} e^{i  q_2 p_2} \ket{q_1 =0,p_2=0},
\ket{\psi(q_1,p_2)} &= \fr{1}{N} e^{i q_2 p_2} \int \mathrm d q_2 \bra{0}  U_{\text{BS}}  \ket{q_1,q_2} \\
&=  \fr{1}{N} \int \mathrm d q_2 e^{i q_2 p_2} \bra{0}  T q_1+ Rq_2,- R q_1 + T q_2 \rangle \nn \\
&=  \fr{1}{N} \int \mathrm d q_2 e^{-\fr{(- R q_1+ T q_2)^2}{2}+i q_2 p_2}  \ket{T q_1+ Rq_2}, \nn
\end{align}
where we have used that the wave function of the vacuum state in the position representation is
\be
\bra{q} 0 \rangle = \fr{e^{-\fr{s^2}{2}}}{\pi^{1/4}},
\ee
and we have absorbed the factor $\pi^{1/4}$ in the normalization constant.
We use now the change of variables 
\be
\fr{T}{R} q = -R q_1+ T q_2,
\ee
from which, using $Tq_1 + \fr{R^2}{T}q_1 +q = q+ \fr{q_1}{T}$, we obtain
\begin{align}
\label{eq:last-step}
\ket{\psi(q_1,p_2)} & = \fr{e^{i \fr{R}{T} q_1 p_2}}{N} \int \mathrm dq e^{- \fr{1}{2} \lt \fr{T}{R} \rt^2 \hat q^2}  e^{+i \fr{p_2}{R} \hat q} \ket{q+ \fr{q_1}{T}} \nn \\
& \propto  e^{i\fr{p_2}{R}  \hat q} e^{-i\fr{q_1}{T}\hat  p}  \int \mathrm dq e^{- \fr{1}{2} \lt \fr{T}{R} \rt^2 \hat q^2} \ket{q}
\end{align}
where we have used $ \ket{q+ \fr{q_1}{T}} = e^{- i \fr{q_1}{T} \hat p} \ket{q}$.
In Eq.(\ref{eq:last-step}) we recognize (apart from an irrelevant global phase factor) a squeezed state $ \hat S(\fr{R}{T} ) \ket{0} \propto \int dq e^{- \fr{1}{2} \lt \fr{T}{R} \rt^2 \hat q^2} \ket{q}$, to which the displacement  $\hat D \lt \fr{1}{\sqrt{2}}\lt \fr{q_1}{T}  + i \fr{p_2}{R}  \rt  \rt \propto e^{i\fr{ p_2}{R}  \hat q} e^{-i\fr{ q_1}{T}  \hat p}$ is applied. This shows that indeed eight-port homodyne detection results in projection onto displaced squeezed states $\hat D(x)\hat S(r) \ket{0}$ with $x= \fr{1}{\sqrt{2}}\lt \fr{q_1}{T}  +i \fr{p_2}{R} \rt$ and $r=  R/T$.

%-------------------------------------------------------------------------------------------------------------------------------------------------------------------------
%\section{Role of the parameter $\phi$}
%\label{app:phi}

\bibliographystyle{apsrev}
\bibliography{bibliography}

\end{document}